\documentclass[sigconf]{acmart}
\AtBeginDocument{%
  }

\copyrightyear{2025}
\acmYear{2025}
\setcopyright{acmlicensed}\acmConference[MM '25]{Proceedings of the 33rd ACM International Conference on Multimedia}{October 27--31, 2025}{Dublin, Ireland}
\acmBooktitle{Proceedings of the 33rd ACM International Conference on Multimedia (MM '25), October 27--31, 2025, Dublin, Ireland}
\acmDOI{10.1145/3746027.3754755}
\acmISBN{979-8-4007-2035-2/2025/10}
\usepackage{threeparttable}
\usepackage{soul}
\usepackage{arydshln}
\usepackage{multirow}

\begin{document}

\title{InterMind: Doctor-Patient-Family Interactive Depression Assessment Empowered by Large Language Models}

\author{Zhiyuan Zhou}
\email{zhouzhiyuan.hfut@gmail.com}
\orcid{0009-0001-9376-7045}
\affiliation{%
  \institution{Hefei University of Technology}
  \city{Hefei}
  \state{Anhui Province}
  \country{China}
}

\author{Jilong Liu}
\email{liujilong0116@gmail.com}
\orcid{0009-0002-7326-9740}
\affiliation{%
  \institution{Hefei University of Technology}
  \city{Hefei}
  \state{Anhui Province}
  \country{China}}

\author{Sanwang Wang}
\email{sanwangwang@whu.edu.cn}
\orcid{0009-0005-0488-3044}
\affiliation{%
  \institution{Wuhan University}
  \city{Wuhan}
  \state{Hubei Province}
  \country{China}
}

\author{Shijie Hao}
\email{hfut.hsj@gmail.com}
\orcid{0000-0003-3181-1220}
\affiliation{%
 \institution{Hefei University of Technology}
 \city{Hefei}
 \state{Anhui Province}
 \country{China}}

\author{Yanrong Guo}
\orcid{0000-0001-6949-4879}
\authornote{Corresponding author.}
\email{yrguo@hfut.edu.cn}
\affiliation{%
  \institution{Hefei University of Technology}
  \city{Hefei}
  \state{Anhui Province}
  \country{China}
}

\author{Richang Hong}
\email{hongrc.hfut@gmail.com}
\orcid{0000-0001-5461-3986}
\authornotemark[1] 
\affiliation{%
  \institution{Hefei University of Technology}
  \city{Hefei}
  \state{Anhui Province}
  \country{China}}

\renewcommand{\shortauthors}{Zhiyuan Zhou et al.}

\begin{abstract}
  Depression poses significant challenges to patients and healthcare organizations, necessitating efficient assessment methods. Existing paradigms typically focus on a patient-doctor way that overlooks multi-role interactions, such as family involvement in the evaluation and caregiving process. Moreover, current automatic depression detection (ADD) methods usually model depression detection as a classification or regression task, lacking interpretability for the decision-making process. To address these issues, we developed InterMind, a doctor-patient-family interactive depression assessment system empowered by large language models (LLMs). Our system enables patients and families to contribute descriptions, generates assistive diagnostic reports for doctors, and provides actionable insights, improving diagnostic precision and efficiency. To enhance LLMs' performance in psychological counseling and diagnostic interpretability, we integrate retrieval-augmented generation (RAG) and chain-of-thoughts (CoT) techniques for data augmentation, which mitigates the hallucination issue of LLMs in specific scenarios after instruction fine-tuning. Quantitative experiments and professional assessments by clinicians validate the effectiveness of our system.
\end{abstract}




\maketitle
\begin{figure}[t]
  \centering
  \includegraphics[width=\linewidth]{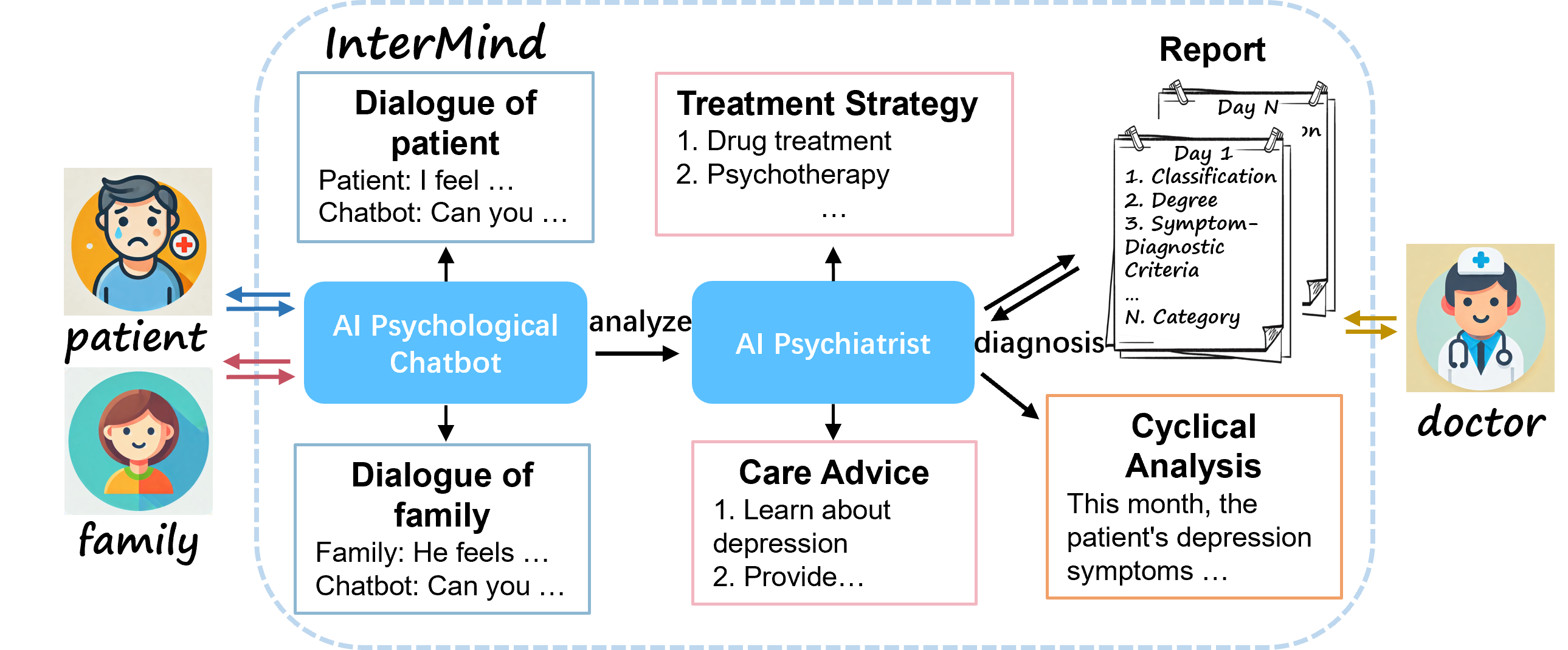}
  \caption{Illustration of proposed LLM-empowered doctor-patient-family interactive depression assessment system.}
  \label{figure1}
\end{figure}
\section{Introduction}
Depression is a serious mental health disorder that significantly affects physical and psychological well-being and can even lead to suicide. According to the World Health Organization (WHO), depression affects over 300 million people worldwide, and untreated mental disorders will account for 13\% of the total disease burden by 2030 \cite{world2017depression}. 
One reason many depression patients struggle to receive evaluation is the lack of an efficient assessment system. Current diagnostic methods rely heavily on self-reports and clinical interviews  \cite{habtamu2023interventions}, which are subjective and often place additional burdens on patients. Moreover, the shortage of psychiatrists cannot keep up with the increasing number of depression patients \cite{johnson2022addressing}, highlighting the need for more efficient assistive diagnostic technologies for doctors.

Although deep learning-based automatic depression detection (ADD) technology \cite{uddin2022deep,niu2021hcag,sadeghi2023exploring,sood2023enhancing, zhou2022hierarchical} has garnered increasing attention from researchers, constructing an efficient depression assessment system still presents several issues:
1) \emph{\ul{Overlooking multi-role engagements in assessment process.}} 
Current depression assessment paradigm, which typically models patients only or in a patient-doctor way \cite{heaukulani2024deploying,kim2024mindfuldiary}, faces challenges due to patients' tendency to inaccurately express or conceal their feelings. Thus, it is crucial to involve the family in the patient's diagnosis and treatment process to provide more objective and comprehensive information, and establish effective interaction among multiple roles in a doctor-patient-family way.
2) \emph{\ul{Lack of in-depth analysis to the results.}} 
Most ADD methods \cite{saggu2022depressnet,wu2022climate,zhang2021depa,nepal2024moodcapture} are limited in classification or regression task with various machine learning or deep learning models, yet the interpretability of the results remains a significant challenge. 
3) \emph{\ul{Fails to align with professional depression assessment scenarios.}} 
Large language models (LLMs) present potential opportunities for assessing depression \cite{xu2024mental,yang2024mentallama,chen2023llm}. However, given the existing limitations of LLMs, such as data scarcity in specific scenarios and hallucinations due to the lack of expertise, these models are insufficient for psychological counseling and assistive diagnostic report generation. For example, LLMs typically generate unstructured content for social media posts to assess depression, but finds that the generated explanations sometimes contain misunderstandings and incorrect reasoning \cite{xu2024mental,yang2024mentallama}. 

To address the above issues, we propose a doctor-patient-family interactive depression assessment system empowered by LLMs, \textbf{InterMind}, as shown in Figure \ref{figure1}. 
This system utilizes the proposed AI Psychological Chatbot and AI Psychiatrist to bridge interactions among doctors, patients, and families.
First, the AI Psychological Chatbot engages with patients in daily conversations, allowing them to describe their experiences and feelings while providing a degree of psychological support during the interaction. The system also enables the AI Psychological Chatbot to concurrently converse with the patient's family members, who can describe the patient's recent condition, such as mood swings or insomnia. By combining dialogues from patients and families, the system helps doctors gain a clearer and more comprehensive understanding of the patient's actual state, reducing the risk of inaccuracies or intentional omissions in the patient's self-reporting.
Subsequently, the AI Psychiatrist is designed to analyze the collected multi-turn dialogues and generate a standardized assistive diagnostic report, which includes the degree of depression, observed symptoms, and relevant diagnostic criteria, rather than simply providing a classification/regression result. Thus, the report offers insights to support the doctor in the diagnostic process and improve diagnostic efficiency. 
Additionally, the AI Psychiatrist provides personalized treatment strategy and care advice for both the patient and their family, aiding in better caregiving and treatment. It can also periodically assess the patient's depression status based on these reports, ensuring continuous monitoring and support.
What's more, our system allows doctors to intervene in the LLM-generated content and enables patients and families to provide feedback on treatment effects, ensuring the accuracy of the assessment and helping doctors adjust treatment strategies timely.

To further improve the capability of LLMs in supporting the roles of AI Psychological Chatbot and AI Psychiatrist in our system, we propose the following methods to enhance their effectiveness in psychological counseling and assistive diagnostic report generation.
For the construction of an AI Psychological Chatbot, we propose the psychological counseling dialogue prompt engineering that leverages diverse experiences from social media depression data to rewrite dialogues. The dialogue fits the psychological counseling scenarios, providing data for instruction fine-tuning, which enhances LLM's ability of psychological counseling. 
For the construction of AI Psychiatrist, we propose diagnostic standard-based report generation which involves a chain-of-thought (CoT) analysis of depression-related symptoms within dialogues. 
Additionally, the retrieval-augmented generation (RAG) technology is applied to match symptoms with corresponding criteria from the DSM-V diagnostic standard \cite{widiger2000toward}. The adopted CoT and RAG techniques enable LLM to generate structured and professional assistive diagnostic reports. 
Based on the above-generated psychological counseling dialogues and their reports, we can build a specialized AI Psychiatrist through instruction fine-tuning.

The contributions of our work are summarized as follows:
\begin{itemize}
\item We propose an LLM-based depression assessment system that introduces a doctor-patient-family paradigm, enhancing caregiving and improving diagnostic efficiency.
\item We use prompt engineering to simulate counseling dialogues for data augmentation, and incorporate CoT and RAG to generate professional diagnostic reports from multisource data. Instruction fine-tuning significantly enhances the LLM's capabilities in psychological counseling and assistive diagnosis.
\item We validate the effectiveness of the proposed AI psychological chatbot and AI Psychiatrist through quantitative experiments and subjective evaluations, demonstrating the efficacy of our approach in building a depression assessment system.
\end{itemize}

\begin{figure*}[h]
  \centering
  \includegraphics[width=\textwidth]{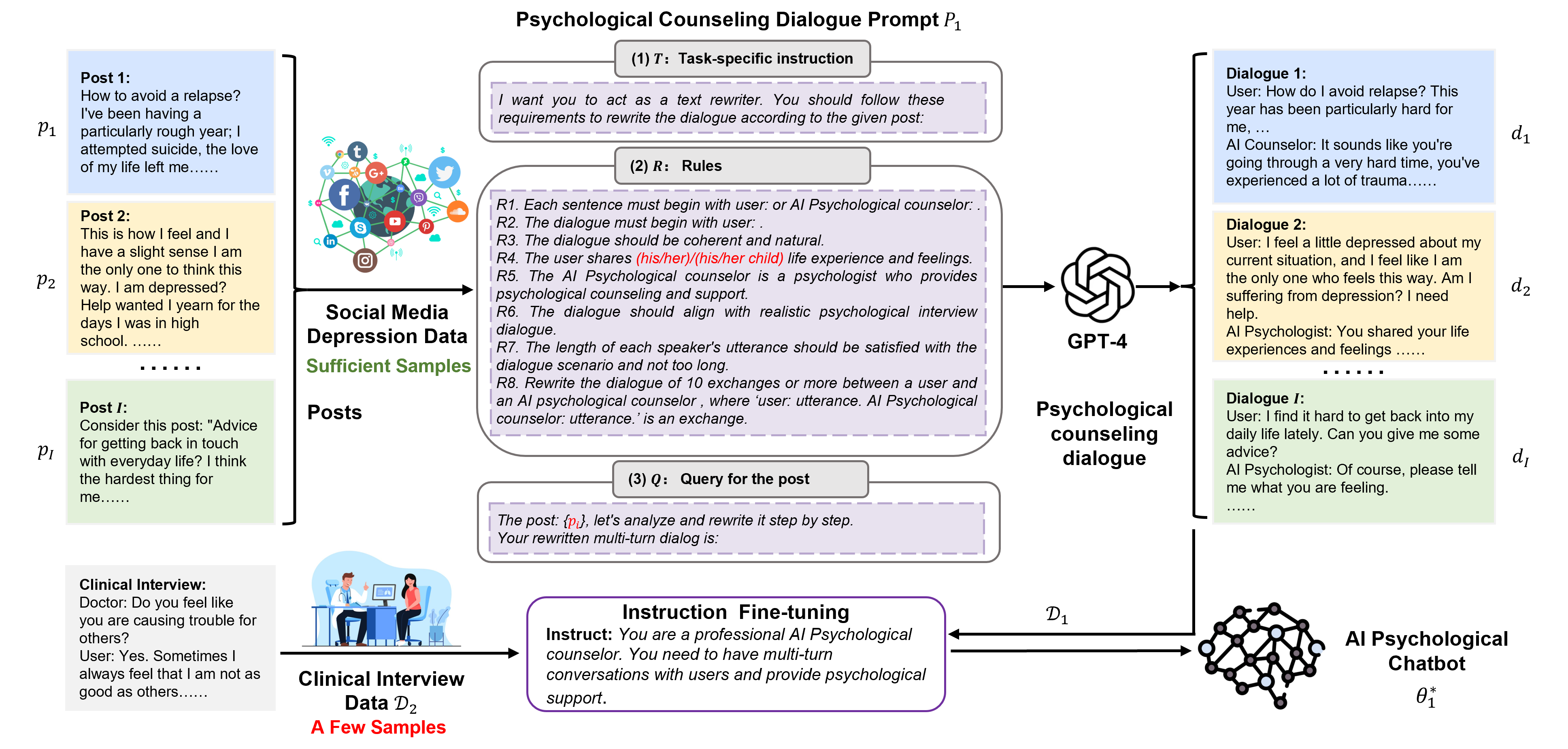}
  \caption{The construction process of AI Psychological Chatbot. The proposed psychological counseling dialogue prompt engineering rewrites the rich experiences and feelings in social media posts \{$p_i$\} into psychological counseling dialogues \{$d_i$\}. The generated dialogues $\mathcal{D}_1$ combined with clinical interview dialogues $\mathcal{D}_2$, form a dataset $\mathcal{D}$ used for instruction fine-tuning LLM, ultimately building an AI Psychological Chatbot ($\theta_1^*$).}
  \label{figure2}
\end{figure*}

\section{Related Works}
\subsection{Mental Health Support Applications}\label{Mental Health Support Applications}
Recent AI-driven mental health applications assist in detection, treatment, and diagnosis. 
MoodCapture \cite{nepal2024moodcapture} uses smartphone images for depression assessment, leveraging facial features while addressing privacy concerns. 
MindfulDiary \cite{kim2024mindfuldiary} is an LLM-based journaling app that helps patients and provides clinicians with insights, validated in a four-week study. 
PsycoLLM \cite{hu2024psycollm} introduces a domain-specific LLM outperforming general models on psychological tasks. However, most existing systems focus on single-role or doctor-patient interactions, overlooking multi-role dynamics.

\subsection{ADD Methods}\label{ADD}
Automatic depression detection (ADD) has gained growing attention, with recent state-of-the-art methods relying on deep learning and large language models (LLMs).
Deep learning-based ADD approaches explore depression patterns through task-specific designs. For example, DeCapsNet \cite{liu2024depression} integrates contrastive learning and symptom capsules aligned with PHQ-9; DEPA \cite{zhang2021depa} introduces self-supervised audio embeddings; and \cite{ijcai2022p725} uses psychiatric scales with a hierarchical attention network based on BERT. While effective, these methods face data scarcity and generalization issues, and often lack interpretability due to their classification/regression framing.
Recently, large language models (LLMs) excel in processing and understanding vast amounts of text data, enabling them to perform a wide range of language-related tasks with high accuracy and versatility across various applications, such as healthcare \cite{Kraljevic2021MedGPTMC} and role-playing \cite{shao2023character}. LLMs also promote the development of ADD methods.
For example, \cite{xu2024mental} shows instruction tuning boosts LLM performance in mental health tasks.
MentaLLaMA \cite{yang2024mentallama} enhances interpretability in social media data.
However, LLMs still struggle with hallucinations and lack domain-specific knowledge in clinical contexts.
In contrast, our approach introduces a doctor-patient-family paradigm, integrating professional diagnostic knowledge and CoT to reduce hallucinations and improve interpretability. It offers a more comprehensive and generalizable diagnostic framework beyond patient- or doctor-centric methods.

\section{Methodology}
\subsection{Overview}
The framework of our InterMind system (Figure \ref{figure1}) is empowered by two designed LLMs components: AI Psychological Chatbot (Sec. \ref{subsec1}) and AI Psychiatrist (Sec. \ref{subsec2}). Specifically, the AI Psychological Chatbot facilitates conversations with patients and their families, allowing them to share their recent experiences and feelings. Subsequently, the AI Psychiatrist integrates the dialogue content from the patient and their family to generate a fixed-format assistive diagnostic report, which involves binary classification, severity degree estimation, related experiences, corresponding diagnostic criteria, and fine-grained categorization. The AI Psychiatrist also provide tailored advice and cyclical analysis to assist with patient treatment, family caregiving, and mental state assessment reference.

\begin{figure*}[h]
  \centering
  \includegraphics[width=\textwidth]{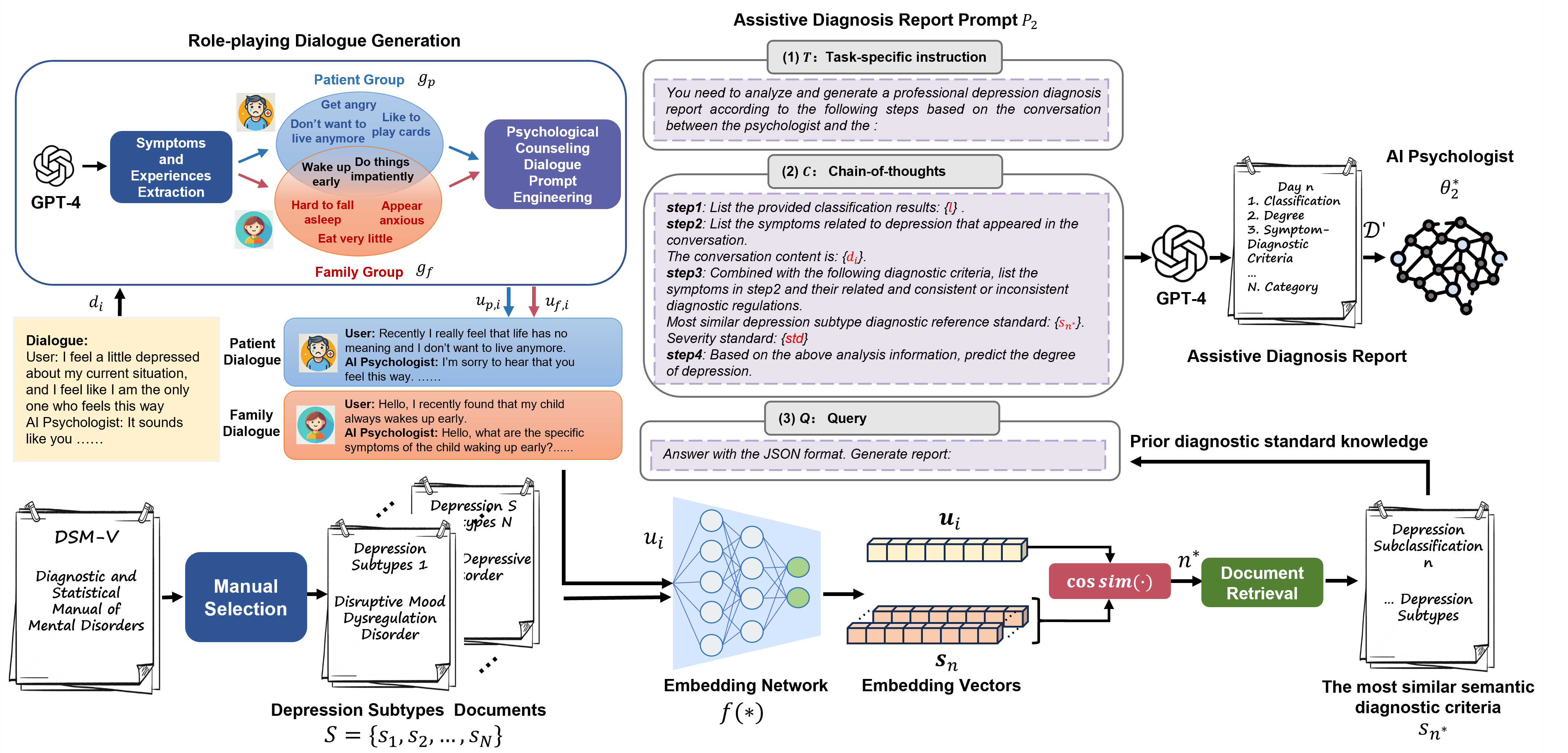}
  \caption{The construction process of AI Psychiatrist. We use RAG techniques to identify the semantically closest section $s_{n^*}$ of the DSM-V criteria $S$ to the user's content $u_i$. Then, through our designed assistive diagnostic report prompt engineering, we apply the CoT approach to generate high-quality assistive diagnostic reports $\{y_i\}$ in a fixed format. These reports $\{y_i\}$, combined with dialogue data $\{d_i\}$, are then used for instruction fine-tuning to build the AI Psychiatrist ($\theta_2^*$).}
  \label{figure3}
\end{figure*}
\subsection{AI Psychological Chatbot}\label{subsec1}
The current use of LLMs to build AI psychological chatbots faces several challenges. On one hand, existing LLMs often struggle to meet the conversational demands in psychological counseling scenarios, frequently generating responses that are either redundant or irrelevant to the context. On the other hand, although instruction fine-tuning can enhance an LLM's performance in specific scenarios, psychological counseling data is difficult to obtain, with public datasets typically containing only a few dozen or hundred examples, which are insufficient for effective LLM training.

Unlike public interview datasets, depression-related social media datasets offers  more abundant experiences and emotions. Therefore, as shown in Figure \ref{figure2}, we develop psychological counseling dialogue prompt engineering, which utilizes designed prompts that guide GPT-4 \cite{achiam2023gpt} to rewrite social media posts into dialogues appropriate for psychological counseling scenarios.
Specifically, inspired by \cite{qiu2023smile} which extends public single-turn dialogues into multi-turn ones, we refine the prompt to make it more standardized, and incorporate more fine-grained psychological counseling scenario settings.  
Our dialogue generation prompt $P_1$ consists of three parts:
\begin{equation}
    P_1=T+R+Q(p_i)
\end{equation}
where instruction $T$ is a description of a specific task, requiring the LLM to follow the rules $R$ to rewrite the post $p_i$ as a dialog. Among the specified eight rules, rules 1-2 ($R1-R2$) control the format of the dialogue's beginning. Rules 3-6 ($R3-R6$) require different characters to share content according to their roles and define the style of the dialogue's setting. Rules 7-8 ($R7-R8$) set limitations on the length of the dialogue and the number of turns. Finally, we provide the post $p_i$ in the query, require the LLM to analyze it step by step, and rewrite the content as a dialogue. Based on the designed prompt $P_1$, the LLM can simulate multi-turn dialogue between the user and an AI counselor. The content derives from authentic experiences and feelings shared in social media posts, thus avoiding fabrications. 

Therefore, we can combine dialogue data $\mathcal{D}_1$ generated from social media with existing clinical interview data $\mathcal{D}_2$ to create a comprehensive multi-turn dialogue dataset $\mathcal{D}=\{\mathcal{D}_1, \mathcal{D}_2\}$. This dataset contains sufficient samples to fine-tune the LLM, enabling the development of an AI Psychological Chatbot. The dataset $\mathcal{D}$ consists of multiple dialogues, where each dialogue $d_i$ contains several turns of interactions $(u_i^j,r_i^j)$, where $u_i^j$ represents the user content in the $j$-th turn of the $i$-th dialogue, and $r_i^j$ represents the corresponding model response. $r_i^j$ consists of a sequence of tokens $\{r_i^j(1),r_i^j(2),$
$...,r_i^j(T_i^j)\}$, where $T_i^j$ is the length of the target sequence. The overall loss function $\mathcal{L}_1(\theta_1; \mathcal{D})$ is given by:
\begin{equation}
\mathcal{L}_1(\theta_1; \mathcal{D}) = \sum_{i=1}^{I} \sum_{j=1}^{J} \left[ - \sum_{t=1}^{T_i^j} \log P\left(r_i^j(t) \mid u_i^j, r_i^j(<t); \theta_1\right) \right]
\end{equation}
where $P(r_{i}^{j}(t)|u_{i}^{j},r_{i}^{j}(<t);\theta_1)$ denotes the conditional probability of the target token $r_{i}^{j}(t)$ given the input $u_i^j$ and the preceding tokens $r_{i}^{j}(<t)=\{r_{i}^{j}(1),...,r_{i}^{j}(t-1)\}$ in the sequence. $\theta_1$ indicates model parameters.
By optimizing  $\mathcal{L}_1(\theta_1; \mathcal{D}) $, the LLM learns the dialogue capability in psychological counseling with learned parameters $\theta_1^*$.

\subsection{AI Psychiatrist}\label{subsec2}
The core task of our AI Psychiatrist is to generate assistive diagnostic results based on patient and family dialogues with an AI Psychological Chatbot. Existing methods \cite{xu2024mental,yang2024mentallama} typically use LLMs to classify depression or provide basic analysis from social media posts, but they lack structured representations, multi-role perspectives, and often rely solely on LLMs' inherent knowledge, leading to hallucinations. While \cite{yang2024mentallama} introduces expert explanations, it does not align with standardized diagnostic criteria.
As we know, the gold standard for depression diagnosis is the DSM-V manual \cite{widiger2000toward} However, the full manual is too long to be directly provided to LLMs as they usually have text length limitation. In the following, we address these challenges by introducing role-playing dialogue generation, retrieval-augmented generation (RAG) and chain-of-thoughts (CoT), as shown in Figure \ref{figure3}.

\subsubsection{Role-playing Dialogue Generation}
To simultaneously model patient and family involvement, we firstly instruct GPT-4 to extract symptoms and experiences set $g$ from each dialogue $d_i \in \mathcal{D}$. 
Then, we divide the set $g$ into two overlapping subsets: the patient group $g_p$ and the family group $g_f$, based on a predefined overlapping rate $q$. The overlapping rate $q$ determines the proportion of symptoms shared between $g_p$ and $g_f$, simulating real-world redundancy (e.g., low mood, appetite loss) and complementarity (e.g., private physiological status or experiences from the patient, observable signs from the family).
After conducting psychological counseling dialogue prompt engineering on $g_p$ and $g_f$ respectively, we simulate differentiated descriptions from the perspectives of the patient and the family, resulting in their respective dialogues, $u_{p,i}$ and $u_{f,i}$.

\subsubsection{Diagnostic Criteria Knowledge Retrieval}
To leverage the knowledge in the DSM-V for generating professional assistive diagnostic reports, we introduce the retrieval-augmented generation (RAG) technology \cite{lewis2020retrieval}. RAG combines the strengths of retrieval-based and generation-based models, which are widely used in natural language processing (NLP) tasks \cite{asai2023self,jeong2024adaptive,edge2024local}. 
We observe that the DSM-V diagnostic criteria are primarily based on patient symptoms. Therefore, we use the RAG technique to match the user's dialogue content with the diagnostic criteria of each depression subtype in DSM-V in terms of similarity. The most similar semantic diagnostic criteria are provided to GPT-4 to generate the report.

Specifically, as shown in Figure \ref{figure3}, the DSM-V is divided into different documents according to the depression subtypes. Let $S=\{s_1,s_2,...,s_N\}$ represents the subtype documents set, where each document $s_n$ corresponds to the symptom descriptions of a specific depression subtype.
For a given user input $u_i=\{u_{p,i},u_{f,i}\}$, and a document $s_n$, we encode them into high-dimensional feature vectors using an embedding network $f(*)$:
\begin{equation}
\mathbf{u}_i = f(u_i), \quad \mathbf{s}_n = f(s_n) \quad \text{for } i = 1, 2, \dots, I
\end{equation}
where $\mathbf{u}_i \in \mathbb{R}^k$ and $\mathbf{s}_n \in \mathbb{R}^k$ are the resulting feature vectors, and $k$ is the dimension of the embedding space. 
Subsequently, the semantic similarity between the user input 
$\mathbf{u}_i$ and each document $\mathbf{s}_n$ is estimated using the cosine similarity:
\begin{equation}
\text{sim}(\mathbf{u}_i, \mathbf{s}_n) = \frac{\mathbf{u}_i \cdot \mathbf{s}_n}{\|\mathbf{u_i}\| \|\mathbf{s}_n\|}
\end{equation}
The index of the document with the highest similarity score is selected as:
\begin{equation}
n^* = \arg\max_{n} \text{sim}(\mathbf{u}_i, \mathbf{s}_n)
\end{equation}
The retrieved document $s_{n^*}$ is used as the diagnostic knowledge standard for the report generation.

\subsubsection{Augmented Generation with Chain-of-thoughts}
The retrieved document $s_{n^*}$, along with the label and severity standard, are provided as prior knowledge to GPT-4. We leverage this  knowledge combined with prompt engineering to generate professional and accurate reports. Specifically, we design assistive diagnosis report prompt engineering. As shown in Figure \ref{figure3}, the prompt for generating an assistive diagnosis report consists of three parts:
\begin{equation}
P_2=T+C(l,d_i,s_{n^*},std)+Q
\end{equation}
where the instruction $T$ describes the task and requires generating the report following the steps of the proposed CoT. CoT $C$ guides the LLM to break down complex tasks into smaller, manageable steps, leading to more accurate and coherent outputs\cite{zhang2022automatic,wei2022chain,wang2023plan}. Thus, we leverage the concept of CoT by setting up four steps. Steps 1-2 guide the LLM to gradually analyze the emerging depression symptoms by combining the dialogue $d_i$ with the label $l$. Step 3 requires matching the symptoms with the relevant diagnostic criteria from the provided diagnostic standards document $s_{n^*}$ and severity standard $std$. Step 4 instructs the LLM to predict the severity of depression based on the preceding information. In the query $Q$, we require the LLM to output the report in a fixed JSON format.

\subsubsection{Instruction Fine-tuning for AI Psychologist}
The reports generated based on each dialogue can serve as labels for the dialogues, thereby $\mathcal{D}'=\left\{(P_2', d_i, y_i) \right\}_{i=1}^{I}$ facilitating the construction of instruction fine-tuning, where $y_i$ represents assistive diagnosis report:
\begin{equation}
\mathcal{L}_2(\theta_2; \mathcal{D}') = \sum_{i=1}^{I} \left[ - \sum_{t=1}^{T_i'} \log P(y_{i}(t)|P_2',d_i,y_{i}(<t);\theta_2) \right]
\end{equation}
where $P_2'=T+C(d_i)+Q$ represents the instruction obtained by removing the prior knowledge such as label and diagnostic standard from $P_2$, ${T_i}'$ is the length of $y_{i}$, and $P(y_{i}(t)|P_2',d_i,y_{i}(<t);\theta_2)$ denotes the conditional probability of the target token $y_{i}(t)$ given the instruct prompt $P_2'$, dialogue $d_i$, and the preceding tokens $y_{i}(<t)=\{y_{i}(1),...,y_{i}(t-1)\}$ in the sequence. $\theta_2$ indicates model parameters.
Through the instruction fine-tuning, our AI Psychiatrist with learned parameters $\theta_2^*$ generate high-quality assistive diagnosis reports in a fixed JSON format. The generated report contains four attributes: binary classification, severity degree estimation, symptom-standard, and subtype category, which overcome the limitations of traditional ADD methods.

\subsubsection{Tailored Advice and Cyclical Analysis Generation}
For advice generation and cyclical analysis, we rely on the general capabilities of the LLM to provide tailored advice and analysis.
We instruct the AI Psychiatrist to provide treatment strategy and care advice to patients and families respectively, based on the symptoms described in reports. Moreover, we also instruct the AI Psychiatrist to provide a cyclical analysis of the mental state over a period of time.

\section{Experimental Results and Analysis}
In this section, we first introduce datasets and experimental details. 
Then, we perform fine-grained quantitative and qualitative experiments on the generated diagnostic report (AI Psychiatrist) and psychological counseling (AI Psychological Chatbot). 
Finally, an ablation study is conducted to verify the contributions of RAG and CoT in report generation.

\subsection{Datasets}
\begin{table*}[]
\caption{Depression Binary Classification Results}
\label{tbl2}
\centering
\resizebox{0.8\textwidth}{!}{
\begin{threeparttable} 
\begin{tabular}{lcccccccccccc}
\hline
\textbf{Dataset}     & \multicolumn{4}{c}{\textbf{MMDA}}                             & \multicolumn{4}{c}{\textbf{DR}}                                   & \multicolumn{4}{c}{\textbf{DepSeverity}}                          \\
\hline
Model                & ACC            & PRE            & REC            & F1             & ACC            & PRE            & REC            & F1             & ACC            & PRE            & REC            & F1             \\
\hline
Chatglm3-6b\cite{glm2024chatglm}          & 0.551          & 0.857          & 0.375          & 0.522          & 0.693          & 0.929          & 0.644          & 0.761          & 0.611          & 0.609          & 0.189          & 0.289          \\
Llama2-Chinese-7b\cite{touvron2023llama}     & 0.571          & 0.789          & 0.469          & 0.588          & 0.575          & 0.861          & 0.524          & 0.652          & 0.592          & 0.513          & 0.405          & 0.453          \\
Baichuan2-7b\cite{yang2023baichuan}         & 0.551          & 0.625          & \underline{0.781} & 0.694          & 0.716          & 0.770          & \underline{0.893}    & 0.827          & 0.425          & 0.415          & \textbf{0.919} & 0.571   \\
GPT-4 \cite{achiam2023gpt}               & 0.592          & \underline{0.875}    & 0.438          & 0.583          & 0.773          & \textbf{0.972} & 0.724          & 0.830          & 0.648          & 0.683          & 0.283          & 0.400          \\
BiLSTM \cite{shen2022automatic} & \underline{0.745}   & \textbf{0.955}          & 0.633    & \underline{0.762}    & 0.305    & 0.263          & 0.429 & 0.326   & \textbf{0.846} & \textbf{0.973}   & 0.410    & 0.577 \\
Roberta \cite{poswiata2022opi}& 0.632   & 0.694          & \underline{0.781}    & 0.735    & 0.224    & 0.225          & 0.367 & 0.279    & \underline{0.746} & 0.478   & 0.379    & 0.423 \\
ChineseMentalBert\cite{ji2022mentalbert} & 0.698    & 0.750          & 0.750    & 0.750    & 0.814    & 0.860          & \textbf{0.899} & 0.879    & 0.702 & 0.688   & 0.507    & 0.584 \\
\hdashline
Ours (Patient)           & 0.694 & 0.774 & 0.750          & \underline{0.762} & \underline{0.832} & 0.919    & 0.853          & \underline{0.885} & 0.719    & \underline{0.773} & 0.472          & \underline{0.586}   \\ 
Ours (Family)           & 0.617 & \underline{0.875} & 0.467          & 0.609 & 0.797 & \underline{0.931}    & 0.791          & 0.856 & 0.590    &0.580 & \underline{0.590}          & 0.500   \\ 
Ours (Both)           & \textbf{0.816} & 0.848 & \textbf{0.875}          & \textbf{0.862} & \textbf{0.840} & 0.929    & 0.853          & \textbf{0.890} & 0.709    & 0.712 & 0.503          & \textbf{0.590}   \\      
\hline                    
\end{tabular}
\begin{tablenotes}[para] 
\footnotesize
\item[1] The bold font and underlined font indicate the best and second-best results, respectively.
\end{tablenotes}
\end{threeparttable}} 
\end{table*}
In this work, we utilize three datasets to evaluate our method: MMDA dataset \cite{jiang2022mmda}, DR dataset \cite{pirina-coltekin-2018-identifying}, and DepSeverity dataset \cite{naseem2022early}. 

\subsubsection{MMDA Dataset }
The MMDA dataset \cite{jiang2022mmda} is a large-scale Chinese resource for mental disorder detection, comprising 524 clinically diagnosed and healthy subjects aged 13–85. Labels are assigned by psychologists using standardized scales, and all samples are collected from face-to-face interviews in controlled, noise-free settings. The Hamilton Depression Rating Scale (HAMD) \cite{ma2021patient} was used to evaluate the degree of depression. Based on the HAMD score, depression is categorized as follows: less than 7 is normal, 7-16 is mild, 17-23 is moderate, and 24 or above is severe. Thus, the threshold score of 7 serves as the boundary between depression and non-depression in the binary classification.

\subsubsection{DR dataset}
The DR dataset \cite{pirina-coltekin-2018-identifying} is an English corpus of Reddit posts for depression detection, comprising self-disclosed depression cases and controls from subreddits. It covers diverse topics to enhance generalization, with all users remaining anonymous. The dataset includes 1838 subjects (1003 train, 430 validation, 405 test), with positive samples from users reporting a depression diagnosis and negative samples from unrelated subreddits.

\subsubsection{DepSeverity dataset}
The DepSeverity dataset\footnote{\url{https://github.com/usmaann/Depression_Severity_Dataset}} \cite{naseem2022early} is a collection of 3553 Reddit posts labeled into four depression severity levels (Minimal, Mild, Moderate, Severe) based on clinical standards (e.g., Beck’s Depression Inventory and Depressive Disorder Annotation). The annotations were validated by multiple annotators via majority voting. The dataset includes preprocessed English text (e.g., spelling correction, anonymization).

\subsection{Experimental Setup}
As existing public datasets lack family-perspective data, we follow the method in Sec. 3.3.1 to split each sample's symptoms into two overlapping subsets: patient group $g_p$ and family group $g_f$, with an overlap rate $q = 0.8$. This yields patient- and family-perspective dialogues $u_{p,i}$ and $u_{f,i}$. We train models using $\{u_{p,i}, y_i\}$, $\{u_{f,i}, y_i\}$, and $\{u_i, y_i\}$ to represent the patient-only (Ours (Patient)), family-only (Ours (Family)), and multi-role (Ours (Both)) paradigms.
For the embedding network $f(*)$, we used the OllamaEmbeddings\footnote{\href{https://python.langchain.com/v0.2/docs/integrations/text_embedding/ollama/}{https://python.langchain.com/v0.2/docs/integrations/text\_embedding/ollama}}. 
We chose Baichuan2 \cite{yang2023baichuan} as the backbone networks of our AI Psychological Chatbot and AI Psychiatrist, and employed the AdamW optimizer \cite{loshchilov2017decoupled} with Low-Rank Adaptation (LoRA) \cite{hu2022lora} for instruction fine-tuning. 
The learning rate was set to $2 \times 10^{-5}$ with a weight decay of $1 \times 10^{-4}$. We trained the model for 4 epochs using a batch size of 1 per device and gradient accumulation steps of 1, with a maximum sequence length of 2048. Mixed-precision training was enabled, and gradient checkpointing was applied to reduce memory consumption.
To ensure consistent data evaluation, we generated dialogues and reports in Chinese.
For MMDA and DepSeverity, we conduct a 90\%/10\% train-test split. For the DR dataset, we use the official training and test set. We train and evaluate our model on the combined datasets ($\mathcal{D}$ and $\mathcal{D}'$), comprising 4677 training and 809 test subjects. Results are reported separately for each dataset.

\subsection{Assistive Report Generation Competency}
\subsubsection{Depression Binary Classification Results}\label{binary}
As shown in Table \ref{tbl2}, we compare the binary classification results in reports generated by the proposed AI Psychiatrist with LLMs (ChatGLM3-6B \cite{glm2024chatglm}, Llama2-Chinese-7B \cite{touvron2023llama}, Baichuan2-7B \cite{yang2023baichuan}, and GPT-4) and supervised models (BiLSTM \cite{shen2022automatic},RoBerta\cite{poswiata2022opi}, and ChineseMentalBert \footnote{\href{https://github.com/zwzzzQAQ/Chinese-MentalBERT}{https://github.com/zwzzzQAQ/Chinese-MentalBERT}}\cite{ji2022mentalbert}) which show strong ability in depression detection. 

From Table \ref{tbl2}, we can see that our method achieves best F1 on the three datasets, which is also superior to the patient-only and the family-only modeling. This phenomenon demonstrates that our method, within the multi-role paradigm involving family dynamics, enables a comprehensive understanding of both patient and family information, ultimately delivering more accurate classification results. 
In addition, the performance of our model is improved significantly compared to Baichuan2-7b across three datasets. This validates that our fine-tuning task can effectively improve backbone model's ability in depression binary classification.
What's more, our method stands out with better performance compared to supervised models which are limited in specific task and lacks interpretability. We also find that GPT-4 tends to be conservative in diagnosing depression, avoiding predicting a subject as depressed without strong evidence. This results in relatively high PRE but low REC, ultimately leading to lower F1 scores.

\subsubsection{Depression Severity Degree Estimation Results}\label{degree}
\begin{table*}[]
\caption{Depression Severity Degree Estimation Results}
\label{tbl3}
\centering
\resizebox{\textwidth}{!}{
\begin{tabular}{c|l|ccc|ccc|ccc|ccc|cc}
\hline
&\textbf{Severity Degree}& \multicolumn{3}{c|}{\textbf{Normal}}       &\multicolumn{3}{c|}{\textbf{Mild}} & \multicolumn{3}{c|}{\textbf{Moderate}} & \multicolumn{3}{c|}{\textbf{Severe}}& \multicolumn{2}{c}{\textbf{Overall}} \\
\hline
Dataset & Model & PRE & REC & F1 & PRE & REC & F1 & PRE& REC & F1 & PRE & REC & F1 & ACC & Weighted-F1 \\
\hline
\multirow{6}{*}{MMDA}& Chatglm3-6b & 0.395  & \underline{0.882}  & 0.545  & \underline{0.800} & 0.148 & 0.250 & 0.333 & \underline {0.500}& \underline {0.400} & 0.000 & 0.000  & 0.000& 0.429 & 0.360 \\
&Llama2-Chinese-7b & 0.354  & \textbf{1.000} & 0.523  & 0.000   & 0.000 & 0.000 & \textbf{1.000}& 0.250  & \underline{0.400}  & 0.000   & 0.000   & 0.000  & 0.367 & 0.214 \\
&Baichuan2-7b  & 0.400  & 0.824 & 0.538& \textbf{1.000}  & 0.037   & 0.071 & 0.167 & \underline{0.500} & 0.250& 0.000  & 0.000   & 0.000 & 0.347 & 0.247 \\
&GPT-4 & \textbf{0.786} & 0.647 & \textbf{0.710}  & 0.700  & 0.519   & 0.596 & 0.125 & 0.250  & 0.167 & \underline{0.143}  & \textbf{1.000}   & \underline{0.250}  & 0.551 & \underline{0.593}  \\
&BiLSTM & 0.381 & 0.471 & 0.421 & 0.643 & 0.667  & \underline{0.655} & 0.000 & 0.000  & 0.000 & 0.000  & 0.000   & 0.000  & 0.531 & 0.501  \\
&RoBerta & 0.375 & 0.177 & 0.240 & 0.537 & \underline{0.815}  & 0.647 & 0.000 & 0.000  & 0.000 & 0.000  & 0.000   & 0.000  & 0.510 & 0.440  \\
&ChineseMentalBert & \underline{0.667}  & 0.235 & 0.348 & 0.571 & \textbf{0.889} & \textbf{0.696} & 0.000  & 0.000  & 0.000 & 0.000 & 0.000  & 0.000  & \underline{0.604}  & 0.504\\
\cdashline{2-16}
&Ours (Patient) & 0.667  & 0.353 & 0.462               & 0.593    & 0.593     & 0.593    & 0.000             & 0.000                  & 0.000                  & 0.111                  & \textbf{1.000}                   & 0.200                  & 0.469                   & 0.491 \\  
&Ours (Family) & 0.556  & 0.588 & 0.571               & 0.625    & 0.185     & 0.286    & 0.015            & 0.250                   & 0.077                  & 0.000                   & 0.000                   & 0.000                 & 0.327     & 0.362 \\  
&Ours (Both) & 0.560  & 0.824 & \underline{0.667}               & \underline{0.800}    & 0.444     & 0.571    & \underline{0.571}             & \textbf{1.000}                   & \textbf{0.727}                  & \textbf{0.500}                   & \textbf{1.000}                   & \textbf{0.667}                  & \textbf{0.633}                   & \textbf{0.619} \\                      
\hline     
\multirow{6}{*}{DepSeverity}&Chatglm3-6b & 0.611  & 0.850  & 0.711  & 0.195 & 0.131 & 0.157 & 0.167 & 0.098& 0.123 & 0.000 & 0.000  & 0.000& 0.530 & 0.456 \\
&Llama2-Chinese-7b & 0.590  & \underline{0.981} & 0.737  & 0.273   & 0.049 & 0.083 & 0.000& 0.000  & 0.000 & 0.000   & 0.000   & 0.000  & 0.580 & 0.444 \\
&Baichuan2-7b  & 0.637  & 0.483 & 0.549& 0.223  & \textbf{0.574}   & \textbf{0.321} & 0.057 & 0.049 & 0.053& 0.167  & 0.022   & 0.038 & 0.389 & 0.387 \\
&GPT-4 & \textbf{0.711} & 0.760 & 0.734  & 0.239  & \underline{0.283}   & 0.260 & 0.219 & 0.171  & 0.192 & 0.214  & 0.136   & 0.167  & 0.530 & 0.518  \\
&BiLSTM & 0.582 & \textbf{0.995} & 0.734 & 0.000 & 0.000  & 0.000 & 0.000 & 0.000  & 0.000 & 0.000  & 0.000   & 0.000  & 0.580 & 0.428  \\
&RoBerta & 0.645 & 0.925 & 0.759 & \underline{0.450} & 0.147  & 0.222 & 0.250 & 0.195  & 0.219 & 0.285  & 0.043  & 0.075  & 0.591 & 0.516  \\
&ChineseMentalBert & 0.685  & 0.913 & \underline{0.783} & \textbf{0.714} & 0.082 & 0.147 & \underline{0.320}  & \underline{0.390}  & \textbf{0.352} & \textbf{0.500} & \underline{0.239}  & \textbf{0.324}  & \textbf{0.627}  & \textbf{0.564}\\
\cdashline{2-16}
&Ours (Patient) & \underline{0.703}  & 0.879 & 0.781               & 0.417    & 0.082     & 0.137    & 0.267             & 0.195                  & 0.225                  & 0.241                  & \textbf{0.283}                   & \underline{0.260}                  & 0.586                   & 0.539 \\  
&Ours (Family) & 0.700  & 0.899 & \textbf{0.788}               & 0.200    & 0.017     & 0.031    & 0.229            & \textbf{0.410}                   & \underline{0.294}                  & 0.308                   & 0.087                   & 0.136                 & 0.582     & 0.513 \\  
&Ours (Both)& 0.677  & 0.879 & 0.765               & 0.286    & 0.262     & \underline{0.274}    & \textbf{0.500}             & 0.122                  & 0.196                  & \underline{0.350}                   & 0.152                 & 0.212                 & \underline{0.592}                   & \underline{0.543} \\                      
\hline  
\end{tabular}}
\end{table*}

For estimation of depression severity degree in our generated reports, we instructed Chatglm3-6b, Llama2-Chinese-7b,  and Baichuan2-7b to determine whether the user in the conversation belongs to the categories of normal, mild, moderate, or severe. For supervised models, we fine-tuned BiLSTM, RoBerta, and ChineseMentalBert in a four-class classification setting. Since the DR dataset does not contain fine-grained labels for different degrees of depression, we report the results on the MMDA dataset and the DepSeverity dataset, as shown in Table \ref{tbl3}.

Compared to binary classification, the task of depression severity degree estimation poses a greater challenge to the models. We can observe that most methods perform poorly in recognizing moderate and severe depression. Supervised models like ChineseMentalBert still demonstrate an advantage in this task, while GPT-4 performs best among LLMs. 
However, supervised models still fails to detect moderate and severe depression on MMDA. Although GPT-4 can detect moderate and severe cases to some extent, its performance is weaker than ours.
Our method demonstrates effectiveness in recognizing moderate and severe depression, achieving the best overall performance on MMDA and the second-best performance on DepSeverity. Through error analysis, we find that mild symptoms such as 'crying', 'fear', 'headache', and 'stress' are often difficult to classify accurately, and some non-depressed individuals expressing temporary negative thoughts are mistakenly identified as depressed. This highlights the challenge in distinguishing depression severity.

\subsubsection{Report Quality Evaluation}
This subsection further validates report quality, particularly the interpretable contents in reports, through subjective evaluations by psychiatrists and quantitative experiments. 

\textbf{Clinical Psychiatrist Evaluation} 
We collaborated with 5 clinical psychiatrists, inviting them to evaluate the generated reports. The evaluation rule is described in the following. For each subject, GPT-4 and our method generate a report respectively. The generated reports are provided to psychiatrists who independently scored each item in the reports. Scored items include binary depression classification, severity degree, each symptom-standard agreement, and subtype category. Psychiatrists score 1 point for each item that they consider as correct and 0 point when they consider it as incorrect. It is important to note that the psychiatrists are not provided with the dataset's ground truth. They score the items depending on their knowledge and experience. In addition, the psychiatrists are unaware of the model's name when they are scoring. The scores are normalized based on the number of ratings for each item. The above scores represent psychiatrists' agreement on the contents of generated reports.

\begin{table}[]
\caption{Psychiatrist Evaluation Results}
\label{tbl5}
\centering
\resizebox{\linewidth}{!}{
\begin{threeparttable}
\begin{tabular}{clcccc}
\hline
\textbf{Dataset} &\textbf{Model}& \textbf{Class} & \textbf{Degree}& \textbf{Symptom\_Standard} &\textbf{Category}       \\
\hline
\multirow{2}{*}{MMDA}&GPT-4  & \textbf{0.917}   & \textbf{0.625}   & 0.954   &\textbf{0.771}   \\
&Ours     & 0.771  & \textbf{0.625}  & \textbf{0.966}    & 0.688  \\
\hline
\multirow{2}{*}{DR}&GPT-4     & \textbf{0.973}    & \textbf{0.703}      & 0.936     & \textbf{0.865}      \\
&Ours   & 0.811     & \textbf{0.703}   & \textbf{0.972}  & 0.784       \\
\hline
\multirow{2}{*}{DepSeverity}&GPT-4       & \textbf{1.000}    & 0.750   & 0.951  & \textbf{0.975}    \\
&Ours      & 0.825   & \textbf{0.800}    & \textbf{0.987} & 0.875    \\
\hline
\end{tabular}
\begin{tablenotes}[para]
\footnotesize
\item[1] Class, Degree, Symptom\_Standard, Category indicate each item in our assistive diagnostic report, representing binary depression classification, severity degree, symptom-standard agreement, and subtype category.
\end{tablenotes}
\end{threeparttable}}
\end{table}

Psychiatrists evaluated 129 subjects, including MMDA's test set and 80 randomly sampled from DR and DepSeverity. Our method outperformed GPT-4 in interpretable content (Symptom\_Standard) across all datasets, providing richer and more accurate diagnostic information. It also excelled in estimating depression severity. However, in binary classification, our method and GPT-4 showed opposite results compared to Table \ref{tbl2}. This discrepancy arises because both psychiatrists and GPT-4 tend to classify mild cases as non-depressed, especially when scores are borderline. Our method aligns more closely with the dataset's ground truth, considering HAMD scores of 7 or higher as depression, which is more beneficial for early depression detection.

\begin{table}[]
\centering
\caption{The 5-fold cross-validation results of our trained supervised referee model}
\label{tbl6}
\resizebox{0.5\linewidth}{!}{
\begin{tabular}{cccc}
\hline
Fold    & PRE   & REC   & F1    \\
\hline
1       & 0.935 & 0.911 & 0.923 \\
2       & 0.955 & 0.933 & 0.944 \\
3       & 0.936 & 0.918 & 0.927 \\
4       & 0.949 & 0.942 & 0.946 \\
5       & 0.975 & 0.911 & 0.942 \\
\hline
Overall & 0.950 & 0.923 & 0.936 \\
\hline
\end{tabular}}
\end{table}

\begin{table*}[]
\caption{Ablation Study of Retrieval-augmented Generation (RAG) and Chain-of-thoughts (CoT)}
\label{tbl7}
\centering
\begin{threeparttable} 
\resizebox{0.8\textwidth}{!}{
\begin{tabular}{lcccccccccccc}
\hline
\textbf{Dataset}     & \multicolumn{4}{c}{\textbf{MMDA}}                             & \multicolumn{4}{c}{\textbf{DR}}                                   & \multicolumn{4}{c}{\textbf{DepSeverity}}                          \\
\hline
Method                & ACC            & PRE            & REC            & F1             & ACC            & PRE            & REC            & F1             & ACC            & PRE            & REC            & F1             \\
\hline
GPT-4       & 0.592	&0.875&	0.438	&0.583&	0.773	&\textbf{0.972}&	0.724	&0.830	& 0.648 &	0.683 &	0.283	& 0.400          \\
GPT-4+RAG & 0.612&	\textbf{0.933}&	0.438&	0.596&	0.820 &	0.950 &	0.805	&0.871&	\textbf{0.746}&	\textbf{0.769}&	0.561&	0.648     \\
GPT-4+CoT  & 0.796&	0.775&	\textbf{0.969}&	0.861&	0.849&	0.854&	\textbf{0.967}&	\textbf{0.910}	&0.654	&0.564	&0.743	&0.641  \\
GPT-4+RAG+CoT & \textbf{0.813}&	0.811&	0.938&	\textbf{0.870}&	\textbf{0.855}&	0.860&	\textbf{0.967}&	\textbf{0.910}&	0.669&	0.581&	\textbf{0.750}&	\textbf{0.655}\\
\hline                    
\end{tabular}}
\end{threeparttable} 
\end{table*}

\textbf{Quantitative Experimental Evaluation}
\begin{figure}[t]
  \centering
  \includegraphics[width=\linewidth]{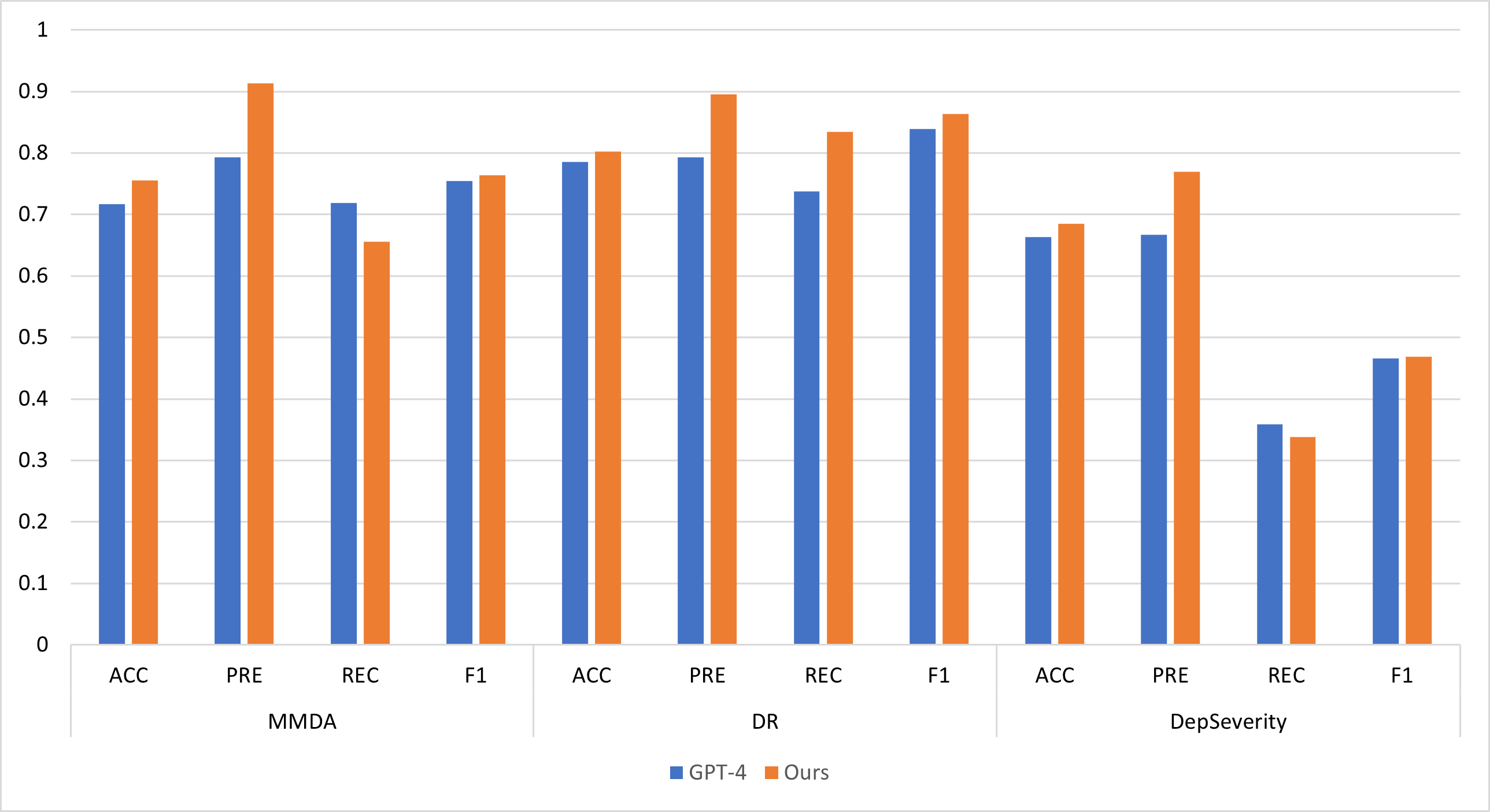}
  \caption{Comparison of the report quality between GPT-4 and our method}
  \label{figure9}
\end{figure}
To further validate the quality of the interpretable content in the reports (Symptom\_\\Standard in Table \ref{tbl5}), we conducted a quantitative experimental evaluation \cite{yang2024mentallama}.
We trained a referee model using ChineseMentalBert to map the symptom and standard sections of reports to their corresponding labels. Table \ref{tbl6} shows the model's performance in evaluating interpretable content through 5-fold cross-validation, demonstrating its strong ability to assess report quality. This referee model is used to evaluate the interpretable content generated by GPT-4 and our method, focusing on the accuracy of binary classification results derived from the content.
Based on the above experimental setting, the reports generated by our method consistently outperform GPT-4 across all three datasets, as shown in Figure \ref{figure9}. While GPT-4 achieves higher recall on the MMDA and DepSeverity datasets, our method attains higher F1 scores and excels in nearly all evaluation metrics. This demonstrates the superior quality of our method's interpretable content, further validating its effectiveness.

\subsection{Psychological Counseling Competency}\label{Counseling}
\begin{figure}[t]
  \centering
  \includegraphics[width=0.9\linewidth]{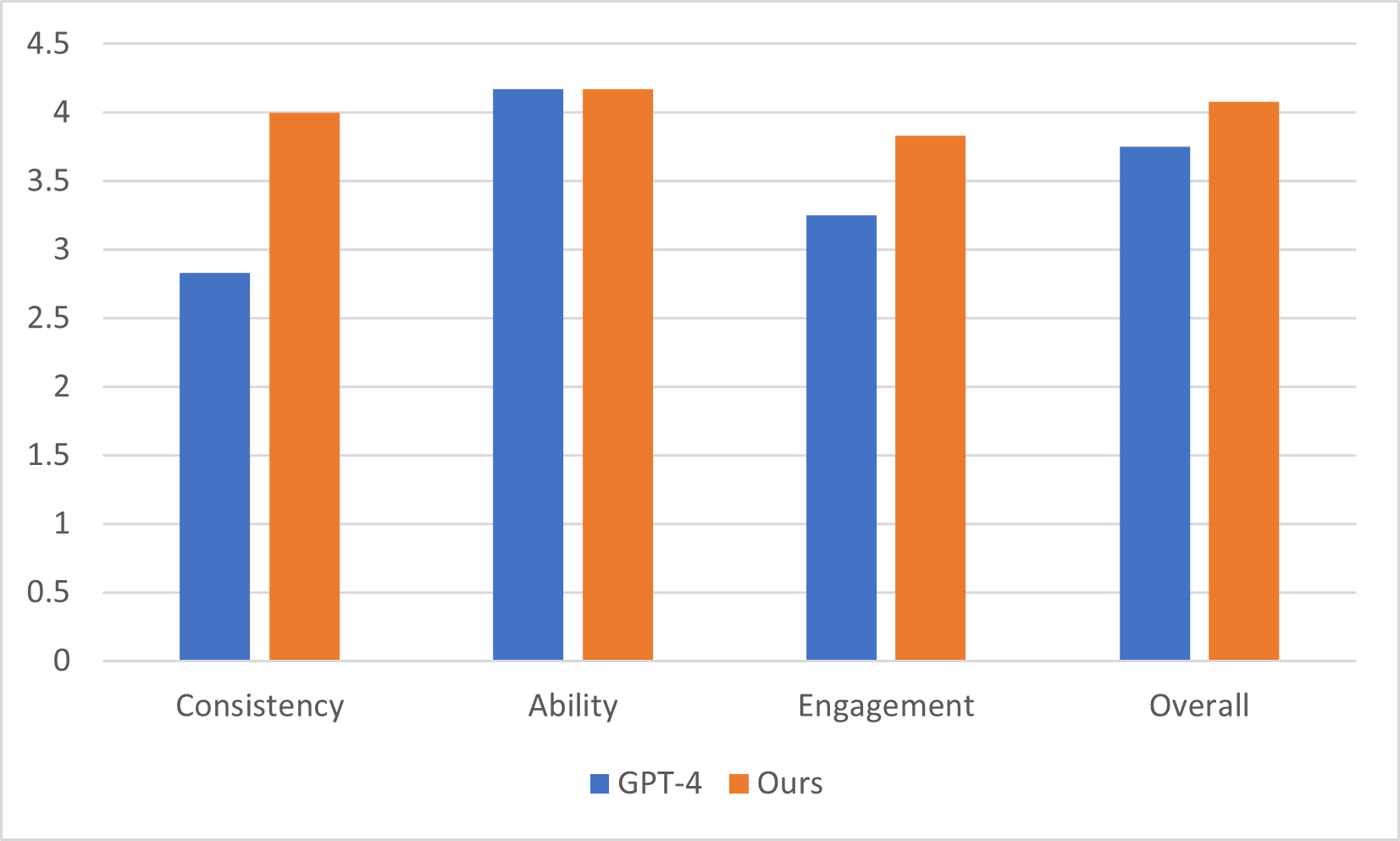}
  \caption{Comparison of the psychological counseling competency between GPT-4 and our method}
  \label{figure10}
\end{figure}

To evaluate the psychological counseling competency of our AI Psychological Chatbot, we invited 18 psychology-trained volunteers to subjectively assess our method and GPT-4. Volunteers interacted with both models and scored them on four metrics: consistency, ability, engagement, and overall performance (each rated 1–5). Consistency measures response alignment with conversational context, ability evaluates counseling competence (e.g., emotional support, symptom inquiry), engagement assesses how well the model encourages user interaction, and overall reflects general performance. Volunteers were kept unaware to the model identities during scoring. 
As shown in Figure \ref{figure10}, our method outperforms GPT-4 in consistency, as GPT-4's formal and lengthy responses are less suited for casual conversations. Both models scored above 4 in ability, demonstrating effective psychological support. Our method achieved higher engagement scores, indicating better encouragement of user interaction, and also surpassed GPT-4 in overall performance.

\subsection{Ablation Study}
To validate the effectiveness of RAG and CoT in GPT-4 for generating assistive diagnostic reports, we conducted an ablation study. Table \ref{tbl7} shows the binary classification results of GPT-4 on three datasets with different combinations of RAG and CoT. Directly instructing GPT-4 yields higher precision but lower recall, reflecting its conservative decision-making. Adding RAG improves performance across all datasets, which addresses GPT-4's previous inability to identify certain depressed subjects. CoT further enhances GPT-4's performance, significantly boosting recall and F1 scores, despite slight precision drops. Combining RAG and CoT (GPT-4+RAG+CoT) achieves the best results, demonstrating that RAG and CoT significantly improve the accuracy of assistive diagnostic report generation.

\section{Conclusion}
In conclusion, we design InterMind, a doctor-patient-family interactive depression assessment system powered by LLMs.It offers psychological support to both patients and families, integrating their information to generate diagnostic reports, treatment plans, and mental state analyses. We enhance LLM performance via prompt engineering for counseling dialogue generation and apply RAG and CoT to produce professional reports, effectively improving the model's capabilities in both counseling and diagnosis. Quantitative experiments and subjective evaluations confirm the system's effectiveness in depression assessment. 
However, this work has several limitations. First, the patient and family perspectives currently rely on LLM-simulated data derived from existing sources, which may not fully reflect real-world situations. Second, although the proposed method aims at enhancing diagnostic efficiency and multi-role interaction in automatic depression assessment, its effectiveness still requires further validation in real clinical settings in collaboration with medical professionals.

\begin{acks}
This work was supported by the National Key Research and Development Program of China under Grant No. 2023YFC2506803, the National Natural Science Foundation of China under Grant No. 62072152, 62172137, 72188101.
\end{acks}

\bibliographystyle{ACM-Reference-Format}
\bibliography{sample-base}

\appendix

\end{document}